*Possible High temperature Superconductivity above 200K mediated by Bose Einstein Condensation of exciton*

Shisheng Lin[1,2]*, Shaoqi Huang,[1] Minhui Yang,[1] Xin Chen[1], Hongjia Bi[1], Kangchen Xiong

[1]College of Information Science and Electronic Engineering, Zhejiang University, Hangzhou, 310027, P. R. China.

[2]State Key Laboratory of Modern Optical Instrumentation, Zhejiang University, Hangzhou, 310027, P. R. China.

*Corresponding author. Email: shishenglin@zju.edu.cn.

**Abstract**: Exciton mediated superconductor is a fascinating quantum phase of matter that occurs when excitons become the dominant excitation in materials, which is also very promising for high temperature superconductor. However, there is no experimental report of exciton mediated superconductivity. Herein, we realize exciton mediated superconductivity from exciton insulator, where the onset transition temperature can reach larger than 200 K. More profoundly, Bose-Einstein condensation (BEC) of exciton can facilitate the formation of exciton insulator and a transition happened from extremely high resistivity of $10^7$ Ω at 153.5 K to $10^0$ Ω at 125 K, indicating a superconducting transition. The resistance of exciton mediated superconductivity can not be absolutely reach zero possibly as a result of drag effect between electrons/holes and excitons. We reveal one rule for exciton mediated superconductivity is that the resistivity is an inverse linear function of the current through the BEC superfluid state, which should be ascribed to the Andreev-Bashkin effect, which reflects the coupling between the BEC state of exciton and Cooper pairs mediated by excitons. Furthermore, a record transition temperature above 200 K has been found for one superconducting sample, which shows the Josephson oscillation dominated by a pendulum-like equation caused by the quantum coupling between superconductor and BEC superfluid of exciton.

The discovery of high-temperature ceramic compound superconductors at 1986 was a groundbreaking event in physics and materials science by Georg Bednorz and Karl Alexander Müller.[1] After that, high temperature superconductor has undergone great experimental development.[2] High temperature superconductors are achieved by doping charge carries into parent compound of Mott insulator, while the mechanism remains unsolved which is the key for understanding and developing novel type of high temperature superconductors. Explaining the high temperature superconductor needs to bring non-phonon possibilities as the Bardeen-Cooper-Schrieffer (BCS) theory utilizing the electron-phonon coupling sets the limits of 40K,[3] which is far below the critical temperature exceeding 77K of liquid nitrogen. Among many attractive mechanisms, the exciton mediated superconductivity is promising for realizing high temperature superconductor as predicted by many scientists including John Bardeen.[4-9] Exciton is the boson binding electron and hole, which has a larger binding energy in low dimensional materials compared with bulk materials.[10,11] The experimental realization of Bose–Einstein condensation (BEC) has been achieved for ultracold atomic vapours under extremely low temperature.[12,13] On the other hand, the exciton based BEC is similar with the superfluid of helium and is attractive as it has many important applications such as polariton laser and quantum computing.[14,15] Actually, there are many reports of possible BEC of the polariton in solid materials detected by optical methods.[4,12] However, there is no claim of exciton mediated superconductivity by BEC as this normally needs a delicately designment of layered structure considering the interaction between electron layer and hole layers. [5]

Basically, exciton mediated superconductivity inherits from the exciton insulator as it transforms doped electrons into Cooper pairs, which is similar with the relationship between Mott insulator and high temperature superconductor. The exciton insulator is a crucial state of matter, the gap of which is related with the binding energy of excitons and their collective ground state, rather than the band structure of the material itself. Instead of electrons and holes moving independently as in standard conductors or insulators, excitons act as composite bosons and can condense into a ground state that spans the entire material. The condensed excitons can exhibit quantum coherence over macroscopic distances, similar to the coherence observed in superfluid and BEC. However, the search for materials that can sustain an exciton insulator phase is challenging, requiring systems with a strong interaction between electrons and holes. Normally, heterostructures made from layering different two-dimensional (2D) materials can exhibit strong excitonic effects due to their unique electronic properties.[16] Layered InAs/GaSb has also been adapted to explore the exciton insulator as the electron and hole have a binding energy larger than energy band gap of InAs or GaSb.[17,18] On the other hand, $Ta_2NiSe_5$ is another material that has attracted attention as a potential exciton insulator.[19] There is rare evidence of exciton superfluidity or superconductivity by electrical measurement up to now.[20,21] Moreover, the drag effect between the excitons may be favored and the superconducting current could be destroyed while the negative or positive current signal will survive.[22-24]

Herein, we realized the bounded exciton insulator. Moreover, a sharp transition from exciton insulator to exciton mediated superconductivity can be repeated and the transition temperature for exciton mediated superconductivity has been lifted up to larger than 200 K. This research will open the avenue of high temperature superconductor through exciton designment.

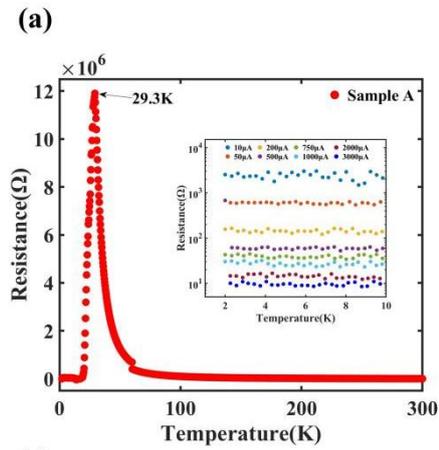
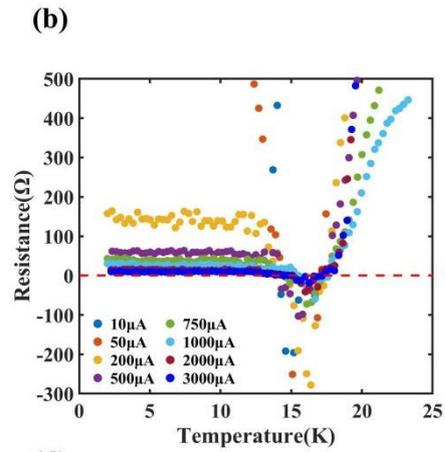
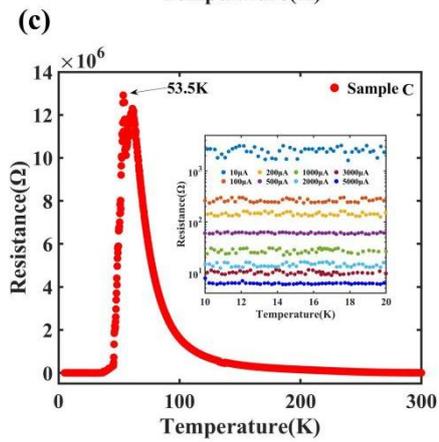
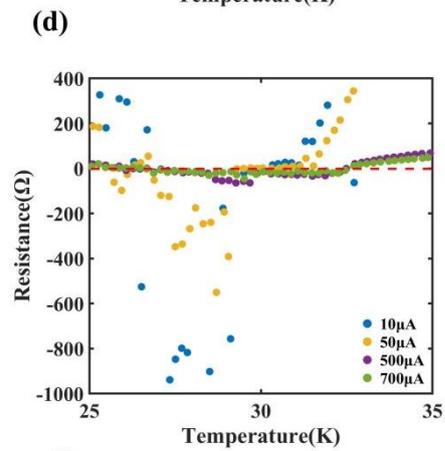
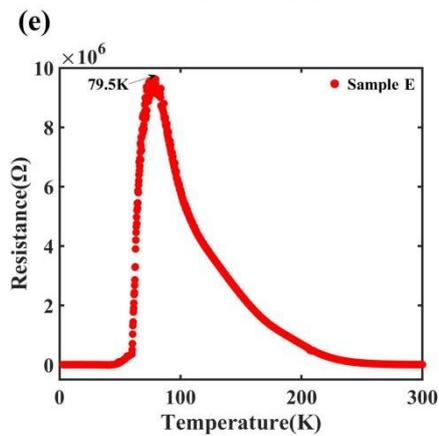
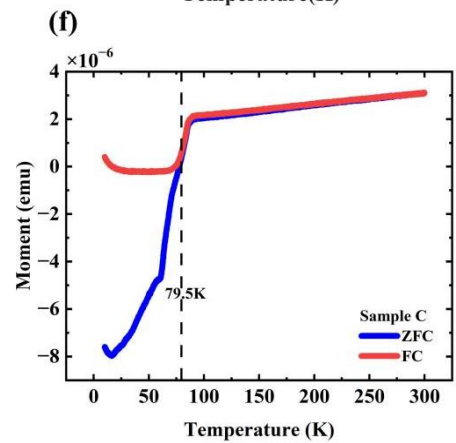
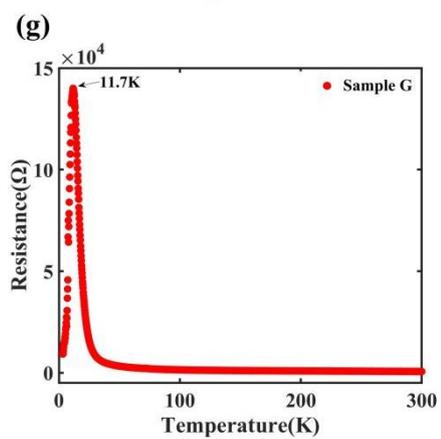
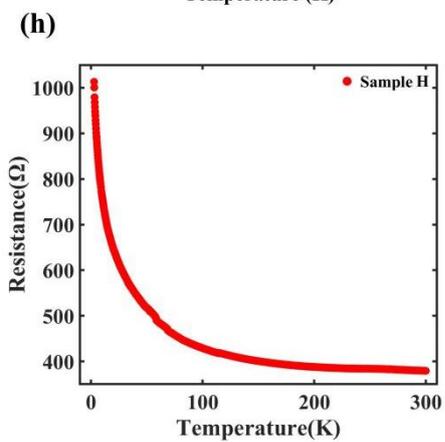

**Fig. 1. Exciton mediated superconductivity and exciton insulator a,** the typical evolution of the resistance as a function of temperature for sample A, where a sharp transition from exciton insulator to superconductivity can be observed; the inset shows the dependence of resistance on the applied current from 2K-10K, **b**, the magnified image of the behavior of exciton mediated superconductivity from 2K-25K, **c,** the transition temperature increases to 53.5K for sample B; the inset shows the dependence of resistance on the applied current from 10K-20K, **d**, the magnified image of the behavior of exciton mediated superconductivity from 25K-35K, **e**, the temperature of the transition from exciton insulator to superconductor increases to 79.5K for sample C, **f**, FC and ZFC curve of the sample E, where an obvious splitting can be observed, **g**, the temperature of the transition from exciton insulator to superconductor increases to 11.7K for sample G, **h**, for comparative sample, no exciton mediated superconductor appears at low temperature.

Figure 1a shows the resistance measurements as a function of temperature, named sample A, where the resistance increase nonlinearly as the temperature decreases and reach an extreme high value of 12 MΩ. By climbing to the highest value, the resistance of sample A suddenly drops down to 2 KΩ at the flowing current of 0.5 μA. It is abnormal that a high resistivity still persists while it suddenly drops more than four orders of magnitudes. Considering there is a drag effect between Cooper pairs and exciton, the residue resistance seems reasonable. The stronger interaction among the excitons, the stronger electromagnetic interaction between Cooper pair and excitons. For reducing such a drag effect, increasing the density of Cooper pairs seems a theoretical solution as the Cooper pairs will screen the strong interaction among the excitons. In the inset of Figure 1a, by increasing the flowing current through the sample at the value of 10μA, 50μA, 200μA, 500μA, 750μA, 1000μA, 2000μA, 3000μA, respectively, the resistance decreases linearly as a function of flowing current. Figure 1b shows the magnified image of the resistance change as a function of temperature from 2K to 25K, where many negative resistance values can be obviously observed. For repeating the growth, we have chosen the similar sample under similar growth condition named sample C. As shown in Figure 1b, the temperature at the highest resistance value of the exciton insulator is around 53.5 K, below which the resistance suddenly drops down to several KΩ with a fluctuating value possibly caused by the quantum drag effect between the Cooper pairs and the excitons. Similarly, by increasing the flowing current through the sample, the resistance decreases at the rate of increased current, which is in accordance with the abovementioned screen effect by the Cooper pairs formed by excitons. Figure 1d also reveal many negative values appear during the fluctuating processes of the resistance, which should be connected with the drag effect between Cooper pairs and excitons. Figure 1e shows another sample with a transition temperature of 79.5 K, where a sharp transition from several MΩ to values crossover zero. As shown in Fig. 1f, there is obvious splitting of the zero-field cooling (ZFC) and field-cooling (FC) curves at the $T_C$ for another sample E grown under similar condition, indicating the superconducting property of the sample E, as the large curvature difference of the two curves below the $T_C$ originates from the large flux pinning force in the FC condition. Figure 1g shows the another sample with a transition temperature of 11.7 K, where a sharp transition from several MΩ to values not reaching zero. For comparative experiment, we show a nonsuperconductive sample with increased resistance as a function of temperature shown in Figure. 1h. However, there is no phase transition of sharp change of resistivity in the whole temperature range, which indicates the sharp transitions observed in Figure 1a, 1c, 1e and 1g are originated from the real electrical signals. We have repeated the measurements in different machines for the reproducibility and the results are satisfied.

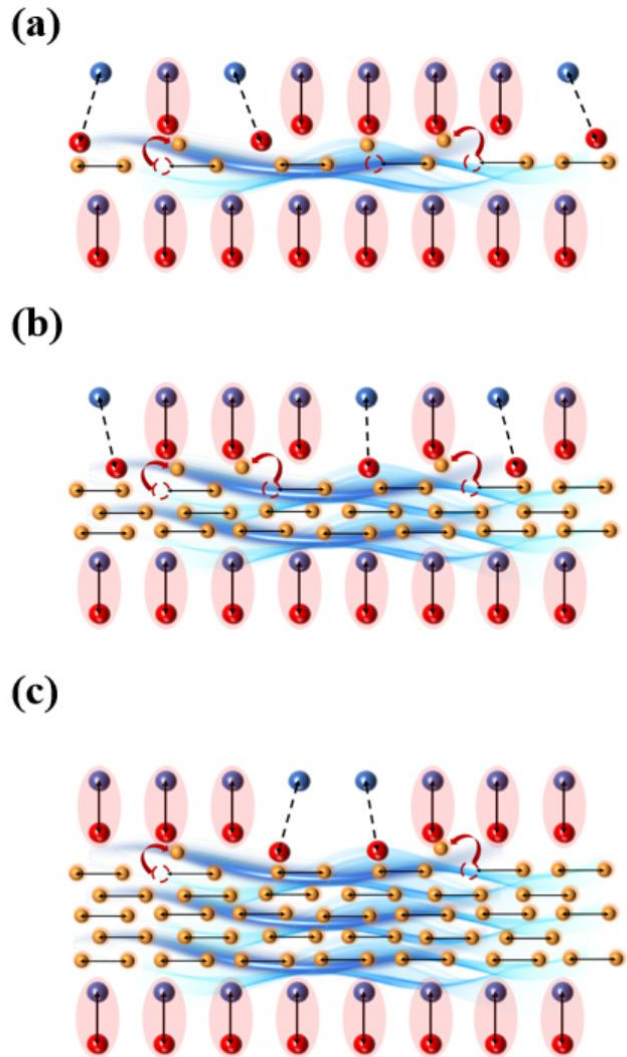

**Fig. 2 Schematic illustration of exciton insulator caused by Bose Einstein condensation.** (a), the interaction between exciton and holes at a small current injection, (b), the interaction between exciton and holes at a medium current injection, (c), the interaction between exciton and holes at a high current injection.

As the holes are confined, permitting the strong interaction between holes with electron centers. After formation of the exciton insulator, the holes injected from the measurement system will be constructed into Cooper pairs by the interaction with the exciton insulator as indicated by many theoretical predictions, where the exciton mediated superconductivity is realized as shown in Figure 1. As shown in Figure 2a, if the injection current is small, the interaction between Cooper pair and exciton insulator is strong, which means the current fluctuation is very severe, in accordance with the experimental results shown in Figure 1. As the flowing current through the exciton insulator increases, the number of Cooper pairs increases, which leads to the screen effect of the interactions among the excitons as shown in Figure 2b. In the case of Figure 2c, further increasing the layer of the density of electrons/holes in the conducting layer will facilitate the formation of more Cooper pairs, which leads to the reduction of resistivity as revealed in Figure 1. The reduced drag effect between exciton and Cooper pairs result in the smaller resistance and the magnitude of current fluctuations, in accordance with the experimental discovery shown in Figure 1.

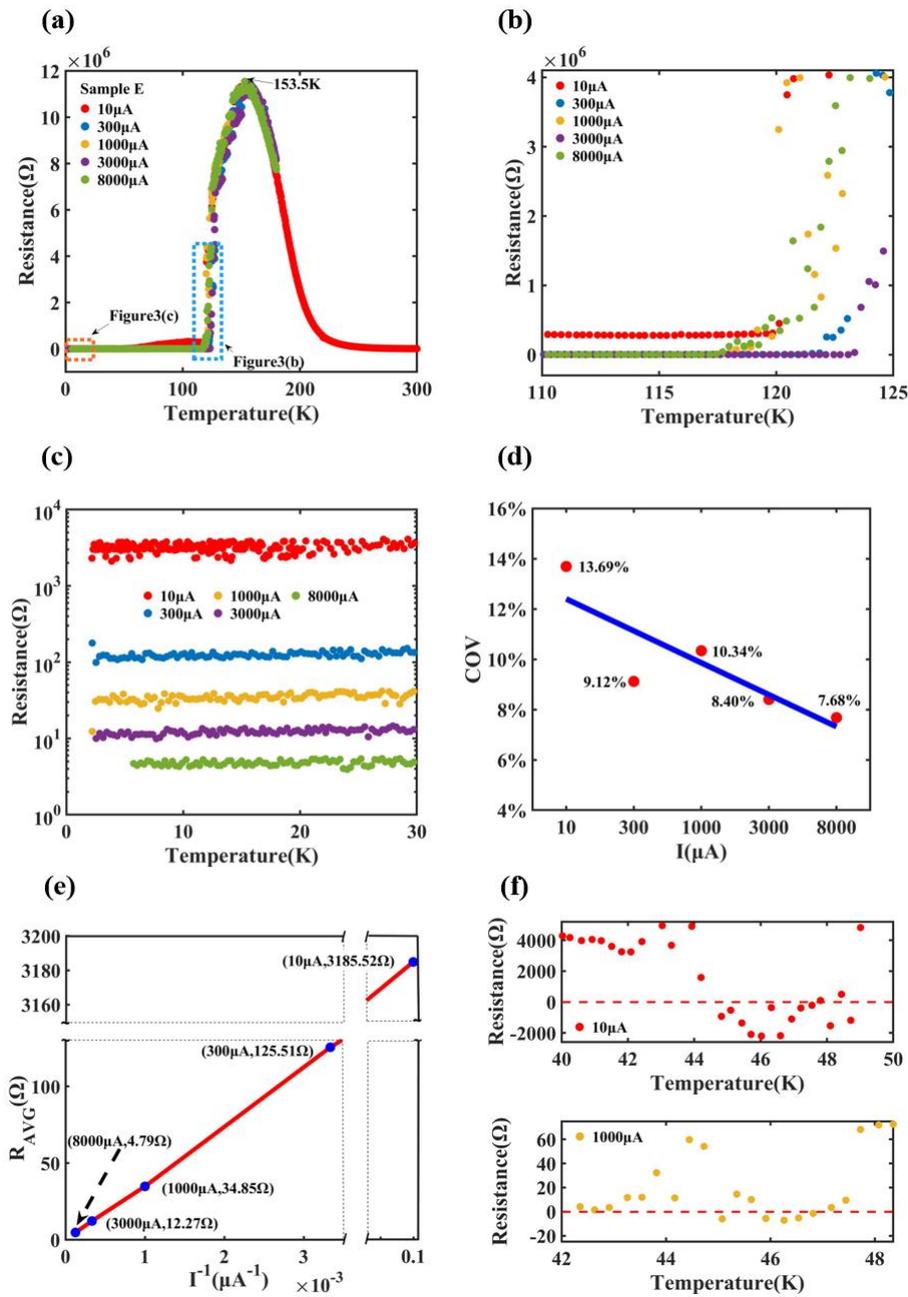

**Fig. 3. High temperature Exciton mediated Superconductivity. a,** the transition from exciton insulator to superconductivity can be observed as high as 153.5K, **b**, the magnified detail of the onset of the exciton superconductivity, **c,** the dependence of resistance as a function of applied current below 30K, where the resistance is inverse proportion to the applied current, **d**, the coefficient of variation of resistance values below 30K with different current, **e**, the dependence of resistance as a function of the reversed current, where a linear function can be clearly clarified, **f**, the resistance crosses over zero near 45 K for applied current of 10 μA and 1000 μA .

Figure 3a shows the appearance of high temperature superconductor, where the transition temperature reaches as high as 153.5 K. The resistance drops down to a stage at the temperature around 120 K shown in Figure 3b. The rules for the resistance as a function of flowing current still apply for the high temperature exciton mediated superconductor. As the flowing current increases from 10μA to 300μA, 1000μA, 3000μA, 8000μA, the resistance decreases linearly from KΩ down to several Ω at the applied current of 8 mA. In accordance with the mechanism mentioned above, the smaller the flowing current through the exciton insulator, the higher magnitude of current fluctuation as shown in Figure 3c. We summarized the coefficient of variation of resistance values (COV) below 30K with different currents, as the current increase, COV shows a downward trend, indicating that the fluctuation of resistance is reduced, as shown in Figure 3d. The linear inversion relationship between resistance and current can be obviously seen in Figure 3e. Moreover, the profound characteristic of negative resistance still appears for this high temperature exciton superconductor, which means some of the electrons or holes are pulled out by the strong interaction between the Cooper pairs and excitons, resulting in a reversed current direction in the method of four probes resistance measurement system.

Figure 3f shows the fluctuation of the resistance at the applied current of 10 μA and 1000 μA. It is found that the current is not kept constant and negative value can be found frequently. The similar behavior can be found in Figure 1b and 1d, where the resistance crosses over zero and fluctuates. On one side, the quantum fluctuation could be dominating the transport processes as the mean field theory is not valid enough for the BEC of excitons.[25] As shown in Figure 2, there are a few Cooper pair flows through the BEC superfluid of exciton, it is inevitable some excitons are broken by the strong electromagnetic interaction between exciton and Cooper pairs. The negative resistance can arise when the superfluid of the BEC of exciton couple with superconducting Cooper pairs, where the counterflow mechanism and unusual current phase relaionship can lead to negative value. The negative resistance under the Andreev-Bashkin effect could be understood through:

$$\Psi_{sf} = |\Psi_{sf}|e^{i\phi_{sf}}, \Psi_{sc} = |\Psi_{sc}|e^{i\phi_{sc}}$$

where $|\psi_{sf}|$ is the magnitude of the exciton BEC superfluid and the $|\psi_{sc}|$ is the magnitude of the superconductor, $\phi_{sf}$ is the coherent phase of the superfluid and $\phi_{sc}$ is the coherent phase of the superconductor. The coupling term between the exciton BEC superfluid and the superconductors is given by: $E_{coupling} = -K_{sf}\cos(\phi_{sf} - \phi_{sc})$, where $K_{sf}$ is the coupling strength between superfluid and superconductors, representing the Andreev-Bashkin coupling as well, which is dependent on the overlap of the wavefunctions at the interface and the interaction energy scale. The induced total energy from the phase gradients of the two components can be expressed as: $E_{total} = K_s \phi_{sf} + K_f \phi_{sc} - E_{coupling}$, where $K_s$ and $K_f$ are the stiffness constants of the respective components. Under specific conditions, the $E_{coupling}$ can dominate, resulting in a current that flows against the applied field, effectively generating negative resistance. On the other hand, similar Andreev-Bashkin effect has been found in 3He-4He system where the superfluid flow of 4He drags the 3He component, leading to the counterflow of current. [26] In our case, the Cooper pairs are superfluid bosons, similar with the BEC of exciton, those two superfluids will interact with each other strongly, leading to the oppositely movement or counterflow of the superfluid components.[27] Counterflow has also been discovered in the electron-hole bilayer structure, and exciton-polariton systems. The negative resistance reflects the interaction at the quantum mechanical level between the two components. The quantum drag and counterflow is crucial for designing device based on superconductivity, superfluidity and excitonic systems.

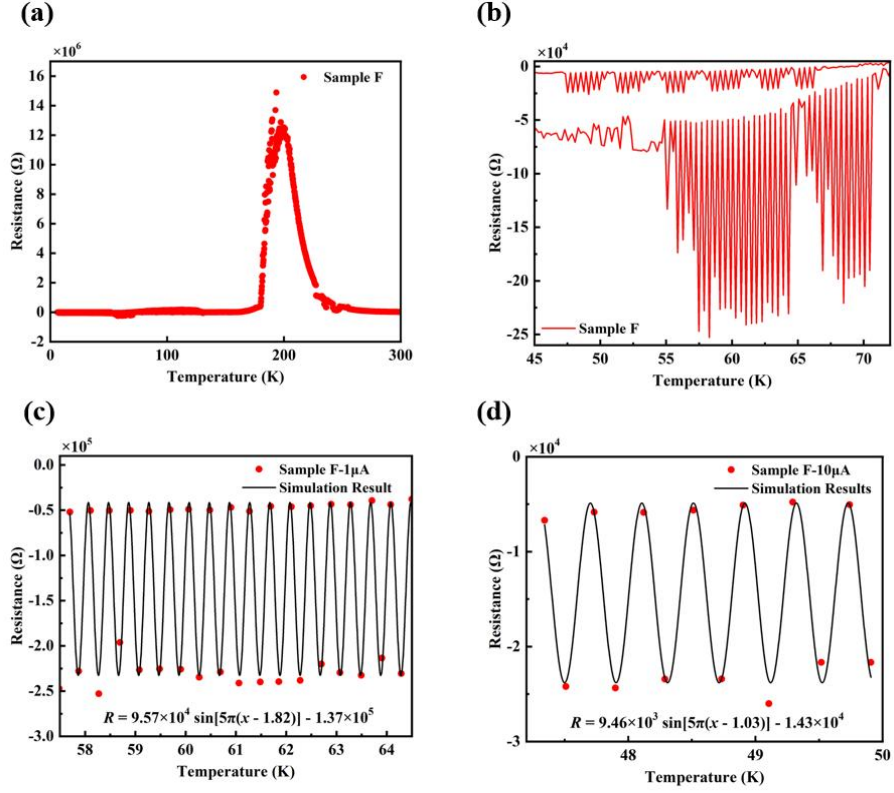

**Fig. 4. Nearly 200K Exciton mediated Superconductor and its phase resonance transport behavior a,** the transition from exciton insulator to superconductivity can be observed as high as 200K, **b**, the oscillation of resistance at the temperature from 45K to 70K, **c,** the simulated and experimental resistance oscillation at the applied current of 1 μA, **d**, the simulated and experimental resistance oscillation at the applied current of 10 μA.

Defining the phase difference between superfluid and superconductor by $\Delta\phi = \phi_{sf} - \phi_{sc}$, the equation for describing such a phase difference can be written by a pendulum-like equation:

$$\hbar \frac{d^2(\Delta\phi)}{dt^2} + K_{sf} \sin(\Delta\phi) = 0$$

Where can be seen that there is a Josephson oscillation proportional to the sin $(\Delta\phi)$ and also there is an exchange of energy between the superfluid.[28-31] Assuming $\Delta\phi \ll 1$, superconductor and superfluid exchange the energy at the resonance frequency. The superfluid and superconductor phases may lock together under strong coupling, stabilizing $\Delta\phi$. Figure 4a shows the temperature dependent resistivity for a sample with a onset transition temperature above 200 K, where a sharp transition from insulating excitonic state to mixed superconducting and superfluid state can be found. As seen from Figure 4b, there is a strong oscillation of the resistance as a function of time as the temperature has a relatively very small effect on the current value at the temperature from 45K to 70K, which is far away from the transition point.[26] The oscillation is much more profound for the applied current of 1 μA compared with that of 10 μA, as the interaction between superconductor and superfluid changes. Considering the electrode distance of 1mm, the coherence length in the superfluid-superconductor system is much longer, further confirming the characteristic of condensate. [32] The simulation results for the Figure 4c and Figure 4d can be written as:

$$R_{1\mu A} = 9.57 \times 10^4 \sin[5\pi(x-1.82)] - 1.37 \times 10^5$$

$$R_{10\mu A} = 9.46 \times 10^3 \sin[5\pi(x-1.03)] - 1.43 \times 10^4$$

where the same frequency can be found, pointing to the same origin of this oscillation as caused by the interaction between superfluid of excitons and superconducting Cooper pairs.[26] It is noteworthy that the temperature parameter is correlated with time point to point and the frequency stay the same for the oscillation under different value of flowing current. This macroscale quantum oscillation will find many intriguing applications in the area of quantum electromagnetic and electronic devices.

In summary, through delicately designing the interaction between excitons and electrons, we have realized the exciton mediated superconductivity at the first time based on the BEC superfluid of excitons. Before the BEC of exciton, the exciton insulator appears accompanying the dramatically increase of the resistance as a function of temperature. We have pointed out the rule for judging the behavior of exciton mediated superconductivity through the current dependent resistance measurements, where the resistance should be inversely linearly with applied current. The high transition temperature around 200 K for the exciton mediated superconductivity has open the numerous chances for designing fascinating quantum devices and systems.

**Experimental Section**

**Superconductive measurements**

The ohmic contact is grown by magnetron sputtering (the spacing between the electrodes was 1.0 mm) for the measurement of low-temperature electrical properties. Superconductivity properties were measured using a low-temperature magnetic field test and specimen preparation system (QUANTUMDESIGN, Dynacool-9) with a test range from 2 K to 300 K. Diamagnetic testing and variable magnetic field resistance testing are performed using a low-temperature magnetic field test and specimen preparation system (QUANTUMDESIGN, MPMS3) with a magnetic field.

**Declaration of Competing Interest**

The authors declare that they have no known competing financial interests or personal relationships that could have appeared to influence the work reported in this paper.

**Authors' Contributions**

S. Lin designed and carried in the experiments, analyzed the data, conceived the study, and wrote the paper. Q. Huang, assisted carrying out the experiments, analyze the data and discuss the results. M. Yang participated the experiments, analyze the data and discuss the mechanism, X. Chen participated the experiments and discuss the data, J. Hong and K. Xiong participated in the experiments. All authors contributed to the preparation of the manuscript.

**Acknowledgements**

S. Lin thanks the support from the National Natural Science Foundation of China (No. 51202216, 51551203, 61774135 and 62474161). S. Lin thanks for the kind discussion with many scientists or colleagues and those discussions are intriguing.